\documentclass[12pt]{elsarticle}

\usepackage{lineno,hyperref}
\usepackage[english]{babel}
\usepackage{amsmath}
\usepackage{amsfonts}
\usepackage{textgreek}
\usepackage{graphicx}
\usepackage{breqn}
\usepackage{undertilde}
\usepackage{appendix}
\usepackage{accents}
\usepackage{enumerate}
\usepackage{caption}
\usepackage{algorithm}
\usepackage{stmaryrd}

\usepackage[noend]{algpseudocode}

\newcommand{\dint}[1]{\ensuremath{\, {\rm d}#1}}
\modulolinenumbers[5]

\journal{arXiv}

\newcommand{\sgn}[1]{\mathrm{sgn}\left(#1\right)}

\newcommand{\ut}[1]{\underaccent{\tilde}{#1}}

\begin{document}
\begin{frontmatter}

\title{Application and adaptation of the truncated Newton method for non-convex plasticity problems.}

\author[TUe]{F.~Bormann}
\author[TUe]{R.H.J.~Peerlings}
\author[TUe]{M.G.D.~Geers}

\address[TUe]{Department of Mechanical Engineering, Eindhoven University of Technology, PO Box 513, 5600 MB Eindhoven, The Netherlands}

\begin{abstract}
	Small-scale plasticity problems are often characterised by different patterning behaviours ranging from macroscopic down to the atomistic scale. In successful models of such complex behaviour, its origin lies within non-convexity of the governing free energy functional. A common approach to solve such non-convex problems numerically is by regularisation through a viscous formulation. This, however, may require the system to be overdamped and hence potentially has a strong impact on the obtained results. To avoid this side-effect, this paper addresses the treatment of the full non-convexity by an appropriate numerical solution algorithm -- the truncated Newton method. 
	The presented method is a double iterative approach which successively generates quadratic approximations of the energy landscape and minimises these by an inner iterative scheme, based on the conjugate gradient method. The inner iterations are terminated when either a sufficient energy decrease is achieved or, to incorporate the treatment of non-convexity, a direction of negative curvature is encountered. If the latter never happens, the method reduces to Newton--Raphson iterations, solved by the conjugate gradient method, with a subsequent line search. However, in the case of a non-convex energy it avoids convergence to a saddle point and adds robustness.
	The stability of the truncated Newton method is demonstrated for the, highly non-convex, Peierls-Nabarro model, solved within a Finite Element framework. A potential drop in efficiency due to an occasional near singular Hessian is remedied by a trust region like extension, which is physically based on the Peierls-Nabarro disregistry profile. The result is an efficient numerical scheme with a high stability that is independent of any regularisation. \par
\end{abstract}


\end{frontmatter}

%
%
\section{Introduction}
%
%
Plastic deformation processes in metals are complex phenomena. What may seem to be a homogeneous material behaviour at a macroscopic scale reveals a significant complexity on lower scales. One complexity relates to non-homogeneous plastic slip, attributed to the motion of dislocations and the formation of dislocation patterns \cite{Ortiz1999}. On the atomistic level, dislocations themselves can be seen as a form of patterning: atomistic slip (disregistry) along a glide plane with a high slip gradient in the dislocation core, as assumed in the Peierls-Nabarro (PN) model \cite{Hirth1982, Joos1994, Schoeck2001}. \par
On all scales, patterning is associated with a non-convex free energy $\Psi$. In (strain gradient) crystal plasticity frameworks \cite{Ortiz1999, Kochmann2011, Miehe2004, Hackl2008, Yalcinkaya2011, Klusemann2013} the non-convexity stems from a latent hardening potential, which is related to the accumulation of trapped dislocations. In models that are based on the mechanics of discrete dislocations or dislocation densities \cite{Giessen1995, Vattre2014, Acharya2001, Fressengeas2011, Hochrainer2014}, the complex interaction between dislocations or dislocation densities gives rise to non-convexity. On the atomistic scale, continuum models establish the disregistry profile of a dislocation through a periodic, and hence highly non-convex, misfit energy that is intrinsic to the glide plane. A modelling approach that can be adopted for all of the above mentioned scales is given by phase-field modelling \cite{Wang2010, Mianroodi2016}, which is equally based on a non-convex potential.\par
%
%
For the solution of common non-linear, but convex, energy minimisation problems the Newton-Raphson method presents a suitable and efficient numerical solution method. Based on the gradient and the Hessian of the total free energy it minimises the energy iteratively until its minimiser is found. For non-convex energy minimisation problems, however, it may fail as an indefinite Hessian $\nabla^2\Psi$ evokes convergence towards a saddle point instead of an energy minimum.\par
One way to deal with the non-convexity is a physical or numerical damping. Adding viscous terms regularises the problem, i.e., it renders the Hessian $\nabla^2\Psi$ positive definite and enables the solution with, e.g., a standard Newton-Raphson scheme. In this respect, two different approaches are commonly followed. i) The non-convex problem is decoupled into a convex elasticity problem and a damped evolution problem of the dislocation (density) configuration. The latter is treated either fully explicitly \cite{Giessen1995, Sandfeld2010} or partially implicitly \cite{Zhang2015}. ii) A damping term is introduced in the non-convex problem and it is solved monolithically with an implicit scheme \cite{Yalcinkaya2011,Bormann2018}.\par
%
%
The introduction of damping, however, may have a significant impact on the results of evolution problems. Particularly in cases where numerical stability requires overdamping or artificial damping, the obtained results may diverge strongly from the solution of the physical problem. For this reason, rate independent (undamped) problems will be considered here. This requires to deal with the non-convexity in the numerical treatment, as a potentially indefinite Hessian $\nabla^2\Psi$ may lead to the breakdown of conventional solution algorithms such as the Newton-Raphson method. \par
%
%
The objective of this paper is hence a robust numerical solution algorithm for non-convex minimisation problems with a performance independent of any regularisation. As a carrier problem the previously proposed Peierls-Nabarro finite element (PN-FE) model is used \cite{Bormann2018}. It considers the idealised plane strain case of a finite domain with a single glide plane for edge dislocations. The glide plane is modelled in alignment with the PN model as a zero-thickness interface that splits the domain into two linear elastic regions. Along the glide plane, any relative tangential displacement, or disregistry, induces a misfit energy based on a periodic, and thus non-convex, potential. As a result, the total free energy, which is comprised of a the linear elastic strain energy and the misfit energy, is non-convex. In earlier research, this non-convexity was dealt with using a viscous regularisation \cite{Bormann2018}, which we aim to avoid here.\par
%
%
Methods that are capable of solving non-convex minimisation problems, although conditionally, are for instance the truncated Newton method (also known as line search Newton-conjugate gradient) \cite{Nash1990,Xie1999,Nash2000,Fasano2013}, BFGS \cite{Li2001,Yuan2018}, the modification of the Hessian to render it positive definite \cite{Schnabel1999a,Kelley2013}, trust region methods \cite{Byrd1987,Conn2000} or Hessian free methods such as the steepest descent \cite{Akaike1959} or the non-linear conjugate gradient method \cite{Wei2008,Yuan2014,Sun2015}. For an overview of these methods the reader is referred to Nocedal and Wright \cite{Nocedal2006}.\par
%
%
In this paper, the truncated Newton method is followed. It poses a stable and efficient solution algorithm for general non-convex problems in which the Hessian $\nabla^2\Psi$ is available at reasonable computational cost. A problem-specific trust region like extension increases its stability and efficiency even further. The adapted truncated Newton method minimises $\Psi$, despite intermittent indefiniteness or near-singularity of $\nabla^2\Psi$, in a straightforward and efficient manner. When close to the minimum, where the energy minimisation problem becomes convex, quadratic convergence behaviour is attained.\par
%
%
This paper is organised as follows. After introducing the non-convex minimisation problem of the PN-FE model in Section \ref{section:Energy_Derivation}, the truncated Newton method is summarised and extended by a trust region like approach in Section \ref{section:Numerical_Method}. In Section \ref{section:Comparison}, the stability and efficiency of the proposed algorithm is demonstrated by comparing it with the standard truncated Newton method and the Newton method with line search for a benchmark problem. The paper concludes with a discussion in Section \ref{section:Discussion}.
%
%
\section{The Peierls-Nabarro finite element model}
\label{section:Energy_Derivation}
\subsection{Problem statement}
The general class of problems considered in this paper is characterised by the two-phase continuum microstructure illustrated in Figure \ref{fig:2-phase-wgp}. It comprises a soft Phase~$\mathrm{A}$ that is separated from the harder Phase~$\mathrm{B}$ by a perfectly and fully coherent phase boundary. Embedded in both phases lies a single glide plane, continuous across the phase boundary. Edge dislocations are introduced into Phase~$\mathrm{A}$ that move as a result of an externally applied deformation towards the phase boundary, where they pile up. When energetically favourable, dislocation transmission into Phase~$\mathrm{B}$ takes place. \par
Notwithstanding the fact that for conciseness the class of problems is limited here to Dirichlet boundary conditions, the derivation for mixed boundary conditions follows a similar approach as outlined below. Equally, an extension towards multiple glide planes or multiple phases is realised straightforwardly.
\subsection{Model formulation}
%
%
Let $R^2$ be the two-dimensional Euclidean space with the global basis vectors \{$\vec{e}_x$, $\vec{e}_y$\} and $\Omega\subset R^2$ the reference configuration of the multi-phase body with outer boundary $\partial \Omega$, consisting of the two phases $\Omega^i\subset\Omega$, $i=\left\{\mathrm{A},\mathrm{B}\right\}$, with outer boundary $\partial\Omega^i\subset\partial\Omega$ and the internal boundary $\Gamma_{\mathrm{pb}}$ separating both phases. The phase boundary orientation is described by its local basis $\{\vec{e}_{n,\mathrm{pb}},\vec{e}_{t,\mathrm{pb}}\}$. The vector $\vec{x}\in\Omega$ is the position vector of any material point in $\Omega$ and $\vec{u}(\vec{x})$ is the displacement vector at $\vec{x}$. Assume further that the deformation characterised by $\vec{u}(\vec{x})$ induces a free energy density consisting of the elastic strain energy density $\psi_{\mathrm{e}}$ and the dislocation induced misfit energy $\psi_{\mathrm{gp}}$ intrinsic to a single, discrete glide plane. In each Phase $i$, the glide plane orientation is characterised by the glide plane normal $\vec{e}_{n,\mathrm{gp}}^{\,i}$ and the glide direction $\vec{e}_{t,\mathrm{gp}}^{\,i}$ which define together the local basis $\{\vec{e}_{t,\mathrm{gp}}^{\,i}, \vec{e}_{n,\mathrm{gp}}^{\,i}\}$. The vector $\vec{s}^{\,i}=s^i\,\vec{e}_{t,\mathrm{gp}}^{\,i}$ denotes the position on glide plane $\Gamma_{\mathrm{gp}}^i$. In the following, the superscripts $i$ for the local basis and for the glide plane position will be dropped for conciseness. \par
\begin{figure}[htbp]
	\centering
	\includegraphics[width=0.8\linewidth]{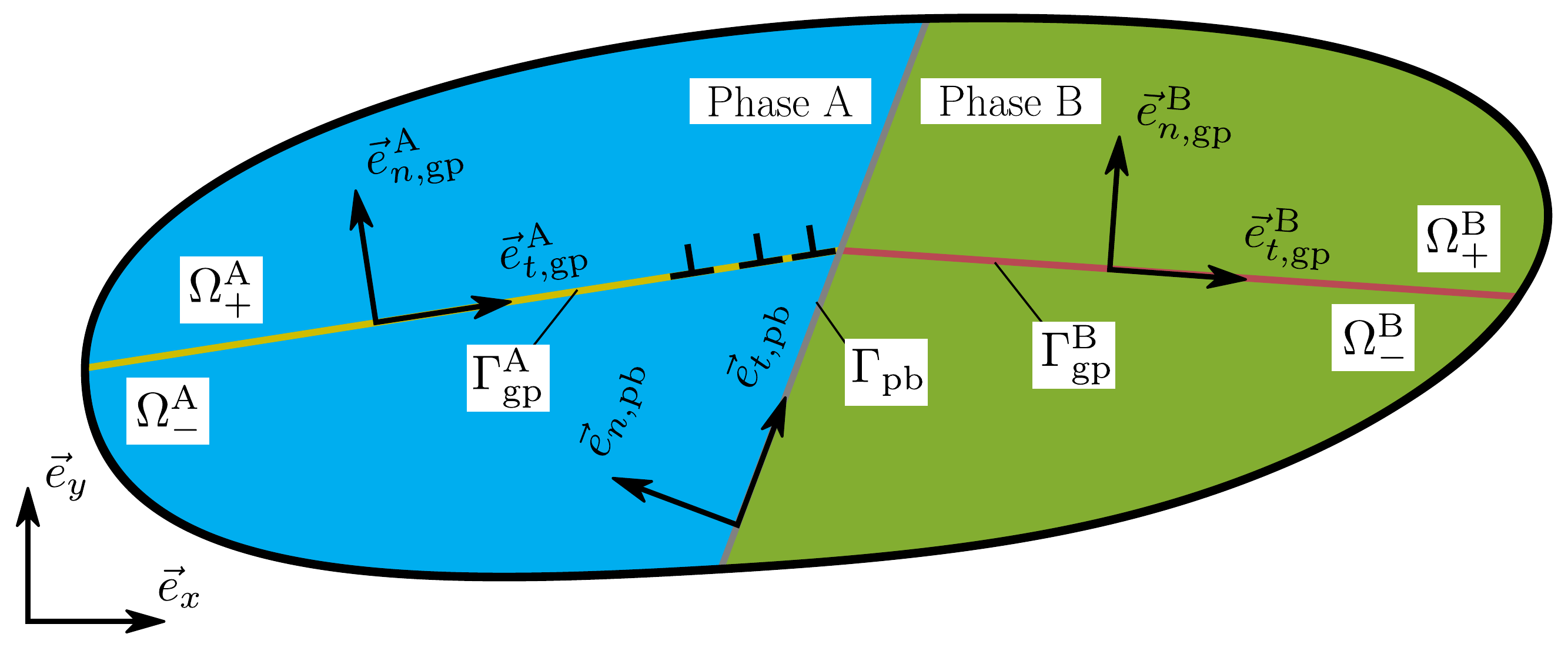}
	\caption{Two-phase microstructure with an internal phase boundary $\Gamma_{\mathrm{pb}}$ and a single glide plane $\Gamma_{\mathrm{gp}}^i$ in each phase.}
	\label{fig:2-phase-wgp}
\end{figure}

Limiting $\psi_{\mathrm{gp}}$ to intrinsic glide plane phenomena, the glide plane can be regarded as an internal boundary $\Gamma_{\mathrm{gp}}^i$ splitting $\Omega^i$ into two elastic subdomains $\Omega_\pm^i\subset\Omega^i$ (see Figure \ref{fig:2-phase-wgp}). $\Omega^i_+$ is defined as the subdomain in the direction of $\vec{e}_{n,\mathrm{gp}}^{\,i}$ and $\Omega^i_-$ the other subdomain. The total free energy $\Psi$ of $\Omega$ can thus be formulated as
\begin{equation}
\label{eq:energy_nanoscale}
\Psi = \int_{\bar{\Omega}} \psi_e\dint\Omega+\int_{\Gamma_{\mathrm{gp}}}\psi_{\mathrm{gp}}\dint\Gamma
\end{equation}
where $\bar{\Omega}=\Omega\setminus(\Gamma_{\mathrm{gp}}\cup\Gamma_{\mathrm{pb}})$ encompasses all subdomains $\Omega_\pm^i$ and $\Gamma_{\mathrm{gp}}=\Gamma_{\mathrm{gp}}^{\mathrm{A}}\cup\Gamma_{\mathrm{gp}}^{\mathrm{B}}$.\par
%
%
Under the assumption of linear elasticity, the strain energy density $\psi_e$ in each subdomain $\Omega_\pm^i$ is defined as
\begin{equation}
\label{eq:energy_elasticity}
\psi_e=\frac{1}{2}\varepsilon^e_{kl}C_{klmn}\varepsilon^e_{mn}
\end{equation}
under a plane strain condition and with the phase-specific fourth-order elasticity tensor $C_{klmn}(\vec{x})=C_{klmn}^i$ for $\vec{x}\in\Omega^i$ and the elastic strain tensor 
\begin{equation}
\label{eq:strain}
\varepsilon^e_{mn}=\frac{1}{2}\left(\frac{\partial u_m}{\partial x_n} + \frac{\partial u_n}{\partial x_m}\right)
\end{equation}
Correspondingly, the linear elastic stress reads
\begin{equation}
\label{eq:stress}
\sigma_{kl} = C_{klmn}\varepsilon^e_{mn}
\end{equation}
\par
%
%
Considering only edge dislocations, the misfit energy $\psi_{\mathrm{gp}}$ is given in accordance with the PN model by \cite{Hirth1982}
%
%
\begin{equation}
\label{eq:energy_PN}
\psi_{\mathrm{gp}} = \gamma_{us}\sin^2\left(\frac{\pi\Delta}{b}\right)
\end{equation}
where $\gamma_{us}$ is the unstable stacking fault energy and $b=b^i$ the magnitude of the Burgers vector. In this paper, $\gamma_{us}$ is approximated by conforming it to linear elasticity in the small strain limit \cite{Hirth1982}, which gives 
\begin{equation}
\gamma_{us} = \frac{\mu b^2}{2\pi^2d}
\end{equation}
with the shear modulus $\mu=\mu^i$ and the interplanar spacing $d=d^{\,i}$. The disregistry $\Delta$ at glide plane position $s$ is defined as
%
%
\begin{equation}
\label{eq:displacement-jump}
\Delta(s) = \vec{e}_{t,\mathrm{gp}}\cdot\left(\vec{u}_{\,+}-\vec{u}_{\,-}\right)
\end{equation}
where $\vec{u}^{\,+}$ and $\vec{u}^{\,-}$ represent the displacement vectors of two initially coincident points on $\Gamma_{\mathrm{gp}}^i$ belonging to $\Omega_+^i$ and $\Omega_-^i$, respectively. The normal opening of $\Gamma_{\mathrm{gp}}^i$ is constrained to be zero:
%
%
\begin{equation}
\label{eq:normal-disregistry}
\vec{e}_{n,\mathrm{gp}}\cdot \left(\vec{u}_{\,+}-\vec{u}_{\,-}\right) = 0
\end{equation}
\par
Along the perfectly bonded phase boundary, displacement and traction continuity are enforced by
\begin{align}
\label{eq:pb-displacement}
\vec{u}^{\mathrm{A}} &= \vec{u}^{\mathrm{B}}\qquad&&\forall\vec{x}\in\Gamma_{\mathrm{pb}}\\
\label{eq:pb-traction}
\boldsymbol{\sigma}^{\mathrm{A}}\cdot\vec{e}_{n,\mathrm{pb}}&=\boldsymbol{\sigma}^{\mathrm{B}}\cdot\vec{e}_{n,\mathrm{pb}}\qquad&&\forall\vec{x}\in\Gamma_{\mathrm{pb}}
\end{align}
On the remote boundary $\partial\Omega$, a displacement field $\vec{u}=\vec{u}_{\mathrm{p}}(\vec{x},t)$, $\vec{x}\in\partial\Omega$, is introduced to load the domain $\Omega$. The dependency of $\vec{u}$ on the pseudo-time $t$ is used to establish the evolution of the system under a gradually increasing load. \par
%
%
\subsection{Finite element discretisation}
To determine the equilibrium configuration of $\Omega$ under the imposed Dirichlet boundary condition at the discrete time $t_n$, the total free energy $\Psi(\partial u_k/\partial x_l,\Delta)$ needs to be minimised with respect to $\partial u_k/\partial x_l$ and $\Delta$. For the solution of this minimisation problem, $\Omega$ is discretised spatially according to the finite element method (FEM). The unknown fields $\vec{u}(\vec{x})$ and $\Delta(s)$ are approximated by
\begin{align}
\label{eq:disc_disp}
\vec{u}^h(\vec{x})&=
\begin{bmatrix}
\vec{e}_x	&	\vec{e}_y
\end{bmatrix}
\underline{N}(\vec{x})\ut{u},
&&\forall\vec{x}\in\Omega_\pm^i\\
\label{eq:disc-Delta}
\Delta^h(s)&= \overline{\underline{N}}(s)\ut{\Delta}&&\forall\vec{x}\in\Gamma_{\mathrm{gp}}^i
\end{align}
where the matrices $\underline{N}(\vec{x})$ and $\overline{\underline{N}}(s)$ contain the shape functions for $\Omega_\pm^i$ and $\Gamma_{\mathrm{gp}}^i$, interpolating the nodal values of $\vec{u}$ and $\Delta$, contained in the column matrices $\ut{u}$ and $\ut{\Delta}$: 
%
%
%
\begin{align}
\underline{N}(\vec{x})=&
\begin{bmatrix}
N_1(\vec{x}) & 0 & N_2(\vec{x}) & 0 & \cdots\\
0 &	N_1(\vec{x}) & 0 & N_2(\vec{x}) &  \cdots
\end{bmatrix}
\\
\overline{\underline{N}}(s)=&
\begin{bmatrix}
\overline{N}_1(s) & \overline{N}_2(s) & \cdots
\end{bmatrix}
\end{align}
%
%
\begin{align}
\ut{u}=&
\begin{bmatrix}
u_{1,x} & u_{1,y} & u_{2,x} & u_{2,y} & \cdots & u_{n,x} & u_{n,y}
\end{bmatrix}
^T\\
\ut{\Delta} = &
\begin{bmatrix}
\Delta_{1} & \Delta_{2} & \cdots & \Delta_{n}
\end{bmatrix}
^T
\end{align}
\par
Following Eq. \eqref{eq:displacement-jump}, it is possible to express $\ut{\Delta}$ in terms of $\ut{u}$ as
%
%
\begin{equation}
\ut{\Delta} = \underline{P}\,\underline{R}\ut{u}
\end{equation}
where the rotation matrix $\underline{R}$ projects the displacements $\ut{u}$, defined in the global basis \{$\vec{e}_x$, $\vec{e}_y$\}, onto the local basis vector $\vec{e}_{t,\mathrm{gp}}$. $\underline{P}$ is established such that Eq. \eqref{eq:displacement-jump} holds, e.g., for a linear interface element
\begin{equation}
\begin{bmatrix}
\Delta^{1-4}\\
\Delta^{2-3}
\end{bmatrix}
=
\begin{bmatrix}
-1	&	0	&	0	&	1\\
0	&	-1	&	1	&	0	
\end{bmatrix}
\begin{bmatrix}
u_{1,t}\\
u_{2,t}\\
u_{3,t}\\
u_{4,t}\\
\end{bmatrix}
\end{equation}
with points 1 and 2 of $\Omega_-$ initially coincident respectively with points 4 and 3 of $\Omega_+$ and $\ut{u}_t=\underline{R}\ut{u}$. Eq. \eqref{eq:disc-Delta} can thus be reformulated in terms of $\ut{u}$ as 
\begin{equation}
\Delta^h(s) = \overline{\underline{N}}\,\underline{P}\,\underline{R}\ut{u}
\end{equation}
\par
Based on Eq. \eqref{eq:disc_disp}, the components of the elastic strain tensor are obtained as
\begin{equation}
\ut{\varepsilon}=
\begin{bmatrix}
\varepsilon_{xx} & \varepsilon_{yy} & \gamma_{xy}
\end{bmatrix}^T
=\underline{B}\ut{u}
\end{equation}
with $\gamma_{xy} = 2\varepsilon_{xy}$ and the standard strain-displacement matrix $\underline{B}$
\begin{equation}
\underline{B} = 
\begin{bmatrix}
\partial N_1/\partial x & 0 & \partial N_2/\partial x & \cdots\\
0 & \partial N_1/\partial y & 0  & \cdots\\
\partial N_1/\partial y & \partial N_1/\partial x & \partial N_2/\partial y & \cdots\\
\end{bmatrix}
\end{equation}
Similarly, the disregistry gradient is defined for later use as
\begin{equation}
\label{eq:Disreg-Grad}
\frac{\mathrm{d} \Delta^h}{\mathrm{d} s}=\overline{\underline{B}}\ut{\Delta}=\overline{\underline{B}}\,\underline{P}\,\underline{R}\ut{u}
\end{equation}
with 
\begin{equation}
\overline{\underline{B}}=
\begin{bmatrix}
\mathrm{d} \overline{N}_1/\mathrm{d}s & \mathrm{d} \overline{N}_2/\mathrm{d} s & \cdots
\end{bmatrix}
\end{equation}
\par
By inserting the discretisation as defined above, the total free energy $\Psi(\partial u_k/\partial x_l,\Delta)$ becomes a function of the nodal displacements $\ut{u}$:
%
%
\begin{align}
\Psi^{\mathrm{h}}(\ut{u})&=\int_{\bar{\Omega}}\psi_e^h(\ut{u})\dint\Omega
+\int_{\Gamma_{\mathrm{gp}}}\psi_{\mathrm{gp}}^h(\ut{u})\dint\Gamma				\nonumber\\
\label{eq:energy_disc}
&=\int_{\bar{\Omega}}\frac{1}{2}\ut{u}^T\underline{B}^T\underline{C}\,\underline{B}\ut{u}\dint\Omega 
+ \int_{\Gamma_{\mathrm{gp}}}\frac{\mu b^2}{2\pi^2d}\sin^2\left(\frac{\pi}{b} \overline{\underline{N}}\,\underline{P}\,\underline{R}\ut{u}\right)\dint\Gamma
\end{align}
%
%
with the Voigt elasticity matrix $\underline{C}$. \par
To incorporate the constraint introduced by the Dirichlet boundary conditions at time $t_n$, the displacement components $\ut{u}$ can be split into the free (i.e., not prescribed) and prescribed displacements $\ut{u}_{\mathrm{f}}$ and $\ut{u}_{\mathrm{p}}$, respectively, with
\begin{equation}
\ut{u} = \underline{G}_f\ut{u}_{\mathrm{f}} + \underline{G}_p\ut{u}_{\mathrm{p}}
\end{equation}
Here, $\underline{G}_f$ and $\underline{G}_p$ are the relevant transformation matrices. The minimisation problem hence reduces to finding, for given $\ut{u}_{\mathrm{p}}$, the local minimiser $\ut{u}_{\mathrm{f}}^*$ of $\Psi^{\mathrm{h}}$ for which $\partial \Psi^{\mathrm{h}}/\partial \ut{u}_{\mathrm{f}}|_{\ut{u}_{\mathrm{f}}^*}=\ut{0}$ and positive definiteness of $\partial^2 \Psi^{\mathrm{h}}/\partial^2 \ut{u}_{\mathrm{f}}|_{\ut{u}_{\mathrm{f}}^*}$ holds.\par
%
%
\subsection{Conventional Newton-Raphson procedure}
The minimisation of $\Psi^{\mathrm{h}}$ poses a nonlinear problem that needs to be solved iteratively. A common way to proceed in solid mechanics is to linearise the stationarity condition $\partial\Psi^{\mathrm{h}}/\partial \ut{u}_{\mathrm{f}}=\ut{0}$ at an initial estimate for the solution and to solve the linearised system to obtain a better estimate. These two steps are repeated until this so-called Newton-Raphson process converges. In terms of the energy, it implies that, in each iteration $i$, $\Psi^{\mathrm{h}}$ is approximated by the second-order Taylor expansion around the current estimate $\ut{u}_{\mathrm{f}}^i$ of $\ut{u}_{\mathrm{f}}^*$:
\begin{equation}
\label{eq:taylor_energy}
\Psi^{\mathrm{h}}\approx \hat{\Psi}^{i+1} :=
\Psi^{\mathrm{h}}(\ut{u}_{\mathrm{f}}^i,\ut{u}_{\mathrm{p}})
+  \delta \ut{u}_{\mathrm{f}}^T\ut{F}^{\,i} 
+ \frac{1}{2}  \delta \ut{u}_{\mathrm{f}}^T \underline{K}^i \delta \ut{u}_{\mathrm{f}}
\end{equation}
with the current gradient $\ut{F}^{\,i}=\ut{F}(\ut{u}_{\mathrm{f}}^{\,i})$, the current Hessian $\underline{K}^i=\underline{K}(\ut{u}_{\mathrm{f}}^{\,i})$ and the update $\delta \ut{u}_{\mathrm{f}} = \ut{u}_{\mathrm{f}}^{i+1}-\ut{u}_{\mathrm{f}}^i$. 
%
%
From Eq. \eqref{eq:energy_disc}, the gradient follows as
%
%

\begin{align}
\label{eq:Quad_Gradient}
\ut{F} = \frac{\partial \Psi^{\mathrm{h}}}{\partial\ut{u}_{\mathrm{f}}}=\left(\frac{\partial \ut{u}}{\partial\ut{u}_{\mathrm{f}}}\right)^T \frac{\partial \Psi^{\mathrm{h}}}{\partial\ut{u}}=
&\underline{G}_f^T\int_{\bar{\Omega}}\underline{B}^T\underline{C}\,\underline{B} \ut{u} \dint\Omega
\nonumber\\
+ 
&\underline{G}_f^T\int_{\Gamma_{\mathrm{gp}}}\underline{R}^T\underline{P}^T\overline{\underline{N}}^T T_t\dint\Gamma
\end{align}
with the glide plane shear traction
\begin{equation}
T_t = \frac{\partial\psi_{\mathrm{gp}}}{\partial\Delta}=\frac{\mu b}{2\pi d}\sin\left(\frac{2\pi}{b} \overline{\underline{N}}\,\underline{P}\,\underline{R}\ut{u}\right)
\end{equation}
Correspondingly, the Hessian reads
%
%
\begin{align}
\underline{K}=\frac{\partial^2 \Psi^{\mathrm{h}}}{\partial{\ut{u}_{\mathrm{f}}}\partial{\ut{u}_{\mathrm{f}}}}
=
&\underline{G}_f^T\int_{\bar{\Omega}}\underline{B}^T\underline{C}\,\underline{B} \dint{\Omega}\,\underline{G}_f\nonumber\\
\label{eq:Quad_Hessian}
+ &\underline{G}_f^T\int_{\Gamma_{\mathrm{gp}}} \underline{R}^T\underline{P}^T\overline{\underline{N}}^T M_{tt}\overline{\underline{N}}\,\underline{P}\,\underline{R} \dint{\Gamma}\,\underline{G}_f
\end{align}
with the glide plane stiffness
\begin{equation}
M_{tt} = \frac{\partial^2\psi_{\mathrm{gp}}}{\partial\Delta^2}=\frac{\mu}{d}\cos\left(\frac{2\pi}{b} \overline{\underline{N}}\,\underline{P}\,\underline{R}\ut{u}\right)
\end{equation}
\par
%
%
As result of the alternating sign of the glide plane stiffness the Hessian $\underline{K}$ may become positive semi-definite, negative semi-definite or indefinite. While the former represents convexity of the quadratic approximation $\hat{\Psi}^{i+1}$, the latter two relate to non-convexity.\par
Assuming convexity, the update $\delta \ut{u}_{\mathrm{f}}$ is obtained by solving the linear system 
\begin{equation}
\label{eq:Newton-Raphson}
\frac{\partial \hat{\Psi}^{i+1}}{\partial \left(\delta \ut{u}_{\mathrm{f}}\right)}
= \ut{F}^{\,i} + \underline{K}^i \delta \ut{u}_{\mathrm{f}} = \ut{0}
\end{equation}
The local minimiser $\ut{u}_{\mathrm{f}}^*$ of $\Psi^{\mathrm{h}}$ (at time $t_n$) is then found by iteratively solving Eq. \eqref{eq:Newton-Raphson} and updating the current estimate $\ut{u}_{\mathrm{f}}^{i+1}=\ut{u}_{\mathrm{f}}^i+\delta \ut{u}_{\mathrm{f}}$ until convergence is achieved with $\ut{u}_{\mathrm{f}}^*\approx\ut{u}_{\mathrm{f}}^{\mathrm{conv}}=\ut{u}_{\mathrm{f}}^{i+1}$ -- i.e., the common Newton Raphson (NR) method. 
For the initial estimate $\ut{u}_{\mathrm{f}}^1$, two choices are adopted in this paper, as follows. i) Given an update of the pseudo-time $t_n$, i.e., an incrementally updated constraint $\ut{u}_{\mathrm{p}}$, the initial estimate is set to $\ut{u}_{\mathrm{f}}^1(t_{k+1})=\ut{u}_{\mathrm{f}}^{\mathrm{conv}}(t_n)$ with $\ut{u}_{\mathrm{f}}^{\mathrm{conv}}(t_0) = \ut{0}$. ii) For the introduction of a dislocation, the converged solution $\ut{u}_{\mathrm{f}}^{\mathrm{conv}}(t_n)$ is perturbed by the approximated displacement field of the new dislocation $\ut{u}_{\mathrm{f}}^{\mathrm{disl}}$. 
The system is then re-equilibrated with $\ut{u}_{\mathrm{f}}^1(t_n) = \ut{u}_{\mathrm{f}}^{\mathrm{conv}}(t_n)+\ut{u}_{\mathrm{f}}^{\mathrm{disl}}$. A detailed description of the dislocation introduction is given in \cite{Bormann2018}.\par
In case of non-convexity, the $\delta \ut{u}_{\mathrm{f}}$ obtained by solving \eqref{eq:Newton-Raphson} does not necessarily indicate the minimum of the quadratic approximation, but possibly a saddle point or even a maximum. Convergence towards $\ut{u}_{\mathrm{f}}^*$ is not guaranteed anymore. Hence, a numerical scheme for non-convex minimisation is required.
%
%
%
%
\section{The truncated Newton method}
\label{section:Numerical_Method}
%
%
A suitable numerical method for the non-convex minimisation problem is provided by the family of truncated Newton methods -- also known as Newton conjugate gradient methods \cite{Nocedal2006,Nash1990,Xie1999,Nash2000,Fasano2013}. They employ a double iterative scheme that 
finds the approximation of $\ut{u}_{\mathrm{f}}^*$ based on the second-order Taylor expansion \eqref{eq:taylor_energy}. 
\subsection{Outer iterations}
\label{sect:outer_it}
In each outer iteration $i$, the current estimate $\ut{u}_{\mathrm{f}}^i$ is updated by means of a suitable outer search direction $\ut{p}^i$ and step length $\alpha^i$:
\begin{equation}
\ut{u}_{\mathrm{f}}^{i+1} = \ut{u}_{\mathrm{f}}^i + \delta \ut{u}_{\mathrm{f}} = \ut{u}_{\mathrm{f}}^i + \alpha^i \ut{p}^i
\end{equation}
The outer search direction $\ut{p}^i$ is determined on the basis of the quadratic potential $\hat{\psi}^{i+1}$ (Eq. \eqref{eq:taylor_energy}) via an iterative scheme, that is outlined in the next section. In case of convexity, $\ut{p}^i$ represents the minimiser of $\hat{\psi}^{i+1}$. With a subsequent line search, that calculates the step length $\alpha^i$, the potential error of $\ut{p}^i$ with respect to the true non-quadratic energy $\psi^h$ is resolved. In this paper, a backtracking line search algorithm with $0<\alpha^i\le 1$ is adopted \cite{Nocedal2006} -- see ahead to Algorithm \ref{alg:Line_Search}. A suitable step length is reached when the Armijo rule is fulfilled:
%
%
\begin{equation}
\label{eq:Armijo}
\Psi^{\mathrm{h}}(\ut{u}_{\mathrm{f}}^{i}+ \alpha^i\ut{p}^i)\le \Psi^{\mathrm{h}}(\ut{u}_{\mathrm{f}}^i)+c\alpha^i\ut{F}^{\,i^T} \ut{p}^i
\end{equation}
where the constant $0<c<1$ is to be chosen small \cite{Nocedal2006}. For the problem considered here, $c=10^{-3}$ is used. Numerical experiments showed a negligible influence on the numerical performance for smaller values of $c$.\par
The outer iterations of the truncated Newton method terminate with $\ut{u}_{\mathrm{f}}^*\approx\ut{u}_{\mathrm{f}}^{\mathrm{conv}}=\ut{u}_{\mathrm{f}}^{i+1}$ when global convergence is achieved. Hereto, the fulfilment of the following inequalities is required in the present study
%
%
\begin{align}
\label{eq:convergence-disp}
\|\delta \ut{u}_{\mathrm{f}}\|_\infty&\le\varepsilon_u b\\
%
%
\label{eq:convergence-force}
\|\ut{F}^{\,i}\|&\le\varepsilon_f\frac{\mu^A b}{2\pi d}b
\end{align}
%
%
where $\varepsilon_u$ and $\varepsilon_f$ are the set tolerances. For Sufficient accuracy of the results, tolerances $\varepsilon_u=\varepsilon_f = 10^{-3}$ are used in the simulations. Note that the unit [force/length] of the reference value in the force convergence criterion \eqref{eq:convergence-force} results from the 2D plane strain formulation.\par
%
%
\subsection{Inner iterations}
Assuming for now convexity of $\hat{\Psi}^{i+1}$, the truncated Newton method establishes the outer search direction $\ut{p}^i$ by approximately solving the Newton equation Eq. \eqref{eq:Newton-Raphson} with the conventional conjugate gradient (CG) method \cite{Nocedal2006} as the inner iterative scheme. The CG method constructs $\ut{p}^i$ by successively minimising $\hat{\Psi}^{i+1}$ in search directions $\ut{d}_j$, $j=1,2,3,\dots$. To avoid confusion with the outer search direction $\ut{p}^i$, $\ut{d}_j$ is referred to as the inner search direction. As the inner iterations $j$ proceed, an increasingly accurate estimate of $\ut{p}^i$ is obtained by the update $\ut{p}_{j+1}^i = \ut{p}_j^i + \kappa_j \ut{d}_j$ where $\kappa_j$ is the CG step length 
\begin{equation}
\kappa_j=\frac{\ut{r}_j^T\ut{r}_j}{\ut{d}_j^T\underline{K}^i\ut{d}_j}
\end{equation}
and $\ut{r}_j$ the residual
\begin{equation}
\ut{r}_{j+1}=\ut{r}_j-\kappa_j\underline{K}^i\ut{d}_j=-\ut{F}^i-\underline{K}^i\ut{p}_{j+1}^i
\end{equation}
The CG method's main feature -- and the reason for its numerical efficiency -- is the property that all $\ut{d}_j$ are conjugate to each other, leading to a fast decrease of $\ut{r}_j$. \par
The inner iterations are terminated with $\ut{p}^i=\ut{p}_j^i$ when a termination criterion is met. Hereto, Nash and Sofer \cite{Nash1990} proposed a test based on the pseudo-potential (cf. \eqref{eq:taylor_energy})
\begin{equation}
Q_j={\ut{p}_j^i}^T\ut{F}^{\,i}+\frac{1}{2}{\ut{p}_j^i}^T \underline{K}^i\ut{p}_j^i
\end{equation}
It is suggested to terminate the CG iterations when
\begin{equation}
\label{eq:CG-convergence}
\frac{\left(Q_j-Q_{j-1}\right)}{Q_j/j}\le\eta^i
\end{equation}
where $\eta^i\in[0,1)$ is the set tolerance. A numerically efficient manner to update $Q_j$ was derived by Fasano and Roma \cite{Fasano2013} as follows
\begin{equation}
Q_j=Q_{j-1}+\left(\frac{1}{2}\sgn{\ut{d}_j^T\underline{K}^i\ut{d}_j}-1\right)\frac{\ut{r}_j^T\ut{r}_j}{\left|\ut{d}_j^T\underline{K}^i\ut{d}_j\right|}
\end{equation}
Ideally, the tolerance $\eta^i$ is chosen adaptively to ensure a loose tolerance distant from the local minimiser $\ut{u}_{\mathrm{f}}^*$ and a tight tolerance in the proximity of the minimiser $\ut{u}_{\mathrm{f}}^*$. In this context, Eisenstat and Walker \cite{Eisenstat1996} proposed 
%
%
\begin{equation}
\label{eq:EW-tolerance}
\eta^{i+1}=\frac{\left\|\ut{F}^{i+1}-\ut{F}^i-\alpha^i \underline{K}^i \ut{p}^i \right\|}{\left\|\ut{F}^i\right\|}
\end{equation}
with $\eta^1\in[0,1)$. Additionally, safeguards are required for a smooth performance of the solution algorithm. A first safeguard prevents a premature small tolerance due to a small step size or a coincidental good agreement between the $\Psi^{\mathrm{h}}$ and its quadratic approximation $\hat{\Psi}^{i+1}$. It requires the pending tolerance $\eta^{i+1}$ to be no less than $(\eta^i)^\zeta$ if the latter is larger than the set threshold $\theta\in[0,1)$. 
The safeguard thus reads
\begin{equation}
\eta^{i+1} \leftarrow \max\left(\eta^{i+1},(\eta^i)^\zeta\right)\quad\mathrm{if}\quad (\eta^i)^\zeta>\theta
\end{equation}
Eisenstat and Walker \cite{Eisenstat1996} proposed $\zeta = (1+\sqrt{5})/2$ based on a convergence of r-order \cite{Ortega1970} $(1+\sqrt{5})/2$ under the assumption of an initial guess $\ut{u}_{\mathrm{f}}^1$ sufficiently close to $\ut{u}_{\mathrm{f}}^*$. This, however, does not hold for the problem considered here; numerical experiments showed that the computational efficiency is strongly increased for $\zeta=1.25$. 
As a second safeguard, an upper bound $\eta_{\mathrm{u}}$ and a lower bound $\eta_{\mathrm{l}}$ are introduced to prevent too loose tolerances and over-solving:
\begin{equation}
\label{eq:EW-bounds}
\eta^{i+1}\leftarrow\max(\min(\eta^{i+1},\eta_{\mathrm{u}}),\eta_{\mathrm{l}})
\end{equation} \par
%
%
\subsection{Truncation}
\subsubsection{Curvature criterion}
To address the potential non-convexity of $\hat{\Psi}^{i+1}$, the truncated Newton method extends the standard CG method by the nonpositive curvature test
\begin{equation}
\label{eq:negative_curvature}
\ut{d}_j^T\underline{K}^i\ut{d}_j\le 0
\end{equation}
As soon as an inner search direction $\ut{d}_j$ meets this criterion, i.e., it points into a direction of negative curvature within $Q^i$, the inner iterations are terminated with $\ut{p}^i=\ut{p}_j^i$, avoiding the direction with negative curvature. The algorithm then proceeds with the backtracking line search, followed by the construction of a new quadratic approximation in a next outer iteration -- see Section \ref{sect:outer_it}.\par
\subsubsection{Disregistry criterion}
Solely criterion \eqref{eq:negative_curvature}, does not necessarily guarantee computational efficiency, as Nocedal and Wright \cite{Nocedal2006} pointed out: a near singular Hessian $\underline{K}^i$ may yield a long and poor outer search direction $\ut{p}^i$, leading to a computationally expensive line search with only a small reduction of $\Psi^{\mathrm{h}}$. 

\begin{figure}[htbp]
	\centering
	\includegraphics[width=0.9\linewidth]{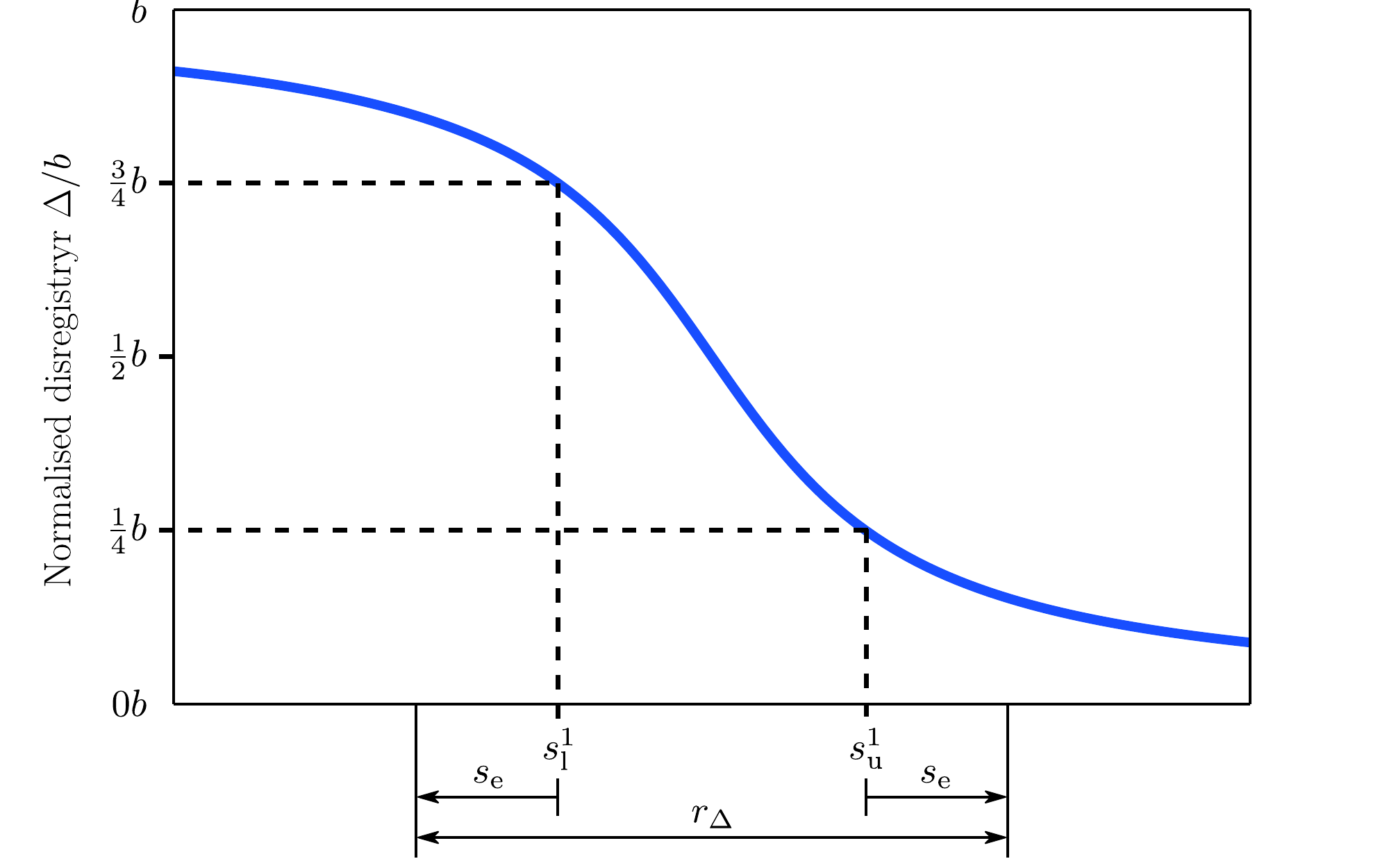}
	\caption{Sketch of the analytical solution for the disregistry profile of a positive PN edge dislocation in an infinite homogeneous medium. $r_{\Delta}$ defines the region in which the disregistry termination criterion is applied.}
	\label{fig:Disreg_ana}
\end{figure}

To prevent this from occurring, an extension of the standard truncated Newton method is proposed by an additional trust region like truncation criterion, as follows. Considering the Peierls-Nabarro dislocation, a strict monotonicity of the disregistry profile is present within the core region: $\mathrm{d}\Delta/\mathrm{d} s<0$ for a positive edge dislocation and  $\mathrm{d} \Delta/\mathrm{d} s>0$ for a negative edge dislocation. This is illustrated in Figure \ref{fig:Disreg_ana}, which sketches the analytical solution of the disregistry profile for a positive PN edge dislocation in an infinite homogeneous medium \cite{Hirth1982}. 
%
%
This requirement leads, using Eq. \eqref{eq:Disreg-Grad}, to the following additional criterion. If for any position $s$ the inequality
\begin{equation}
\label{eq:Disreg_check}
\sgn{\overline{\underline{B}}\,\underline{P}\,\underline{R}\left(\ut{u}^i+\underline{G}_f\ut{p}_{j+1}^i\right)} \neq \sgn{\overline{\underline{B}}\,\underline{P}\,\underline{R}\ut{u}^i}
\end{equation}
yields, i.e., the updated inner search direction $p_{j+1}^i$ leads to a change of the sign of the disregistry gradient, presuming a step length of $\alpha^i = 1$, the inner iteration is truncated with $\ut{p}^i=\ut{p}_j^i$. This criterion is generally violated as a result of the near singularity of $\underline{K}$, which arises through a negative glide plane contribution (cf. Eq. \eqref{eq:Quad_Hessian}). Thus, for computational efficiency, the disregistry criterion is checked only in a region $r_{\Delta}$ around each dislocation, as shown in Figure \ref{fig:Disreg_ana} for a positive dislocation. $r_{\Delta}$ is determined as the region in which $M_{tt}<0$, i.e., $b/4\le\Delta\le3b/4$, extended by a small distance $s_{\mathrm{e}}$. In this paper $s_{\mathrm{e}}=b$ is chosen. Numerical experiments showed that a small increase or reduction of $s_e$ has a negligible influence on the numerical performance. \par
%
%
\subsection{The overall incremental-iterative algorithm}
With this additional criterion, one has to be aware of the impact of the boundary constraints $\ut{u}_{\mathrm{p}}$ applied on $\partial\Omega$. Following Eq. \eqref{eq:Quad_Gradient}, an incremental update of $\ut{u}_{\mathrm{p}}$ induces a rather strong gradient (residual) $\ut{F}^i$ ($-\ut{r}_1$) concentrated on the nodes adjacent to $\partial\Omega$. With an increasing number of inner iterations $j$, the residuals $\ut{r}_j$ spread iteratively inwards into the bulk $\Omega$. In this context, high residuals can emerge at the glide plane and invoke premature termination due to criterion \eqref{eq:Disreg_check}. It is thus essential to split at each pseudo-time $t_n$ the solution process into two steps. The first step freezes the current disregistry profile along the glide plane $\Gamma_{\mathrm{gp}}$ with $\delta\Delta = 0$ and assesses the elastic response emanating from $\ut{u}_{\mathrm{p}}$. It comprises a purely convex minimisation and requires therefore only a single outer iteration $i$ using a tight tolerance $\eta_{\mathrm{ini}}$. In the second step the complete non-convex energy is iteratively minimised until global convergence is attained. \par
The main algorithm of the truncated Newton method, executed at each pseudo-time $t_n$, is outlined in Algorithm \ref{alg:trN-alg} with the Sub-algorithms \ref{alg:Inner_Loop} and \ref{alg:Line_Search}.
\makeatletter
\def\BState{\State\hskip-\ALG@thistlm}
\makeatother

\begin{algorithm}
	\caption{The adapted truncated Newton method}\label{alg:trN-alg}
	\begin{algorithmic}[1]
		\State Set tolerances $\varepsilon_u$, $\varepsilon_f$, $\eta_{\mathrm{ini}}$, $\eta_{\mathrm{u}}$ and $\eta_{\mathrm{l}}$
		\State Initialisation of $\ut{u}_{\mathrm{p}}$, $\ut{u}_{\mathrm{f}}^1$ and gradient $\ut{F}^1$
		\For{$i=1,2,3,\dots$}

		\If {$i=1$}
		\State Constrain disregistry profile with $\delta\ut{\Delta} = \ut{0}$
		\State Set tolerance $\eta^1=\eta_{\mathrm{ini}}$
		\Else
		\State Lift disregistry constraint
		\State Set tolerance $\eta^i$ using Eq. \eqref{eq:EW-tolerance}-\eqref{eq:EW-bounds} and $\eta^2=\eta_{\mathrm{u}}$
		\EndIf

		\State Calculate Hessian $\underline{K}^i$
		%
		%
		\State Calculate $\ut{p}^i$ using function Inner Loop (Algorithm \ref{alg:Inner_Loop})
		%
		%
		\State Calculate $\alpha^i$ using function Backtracking Line Search (Algorithm \ref{alg:Line_Search})
		\State Set $\ut{u}_{\mathrm{f}}^{\,i+1}=\ut{u}_{\mathrm{f}}^{\,i}+\alpha^i\ut{p}^i$ 
		\State Update gradient $\ut{F}^{i+1}$
		\State Convergence test using Eq. \eqref{eq:convergence-disp} and \eqref{eq:convergence-force}
		\If{Convergence}
		\textbf{exit} with $\ut{u}_{\mathrm{f}}^*\approx\ut{u}_{\mathrm{f}}^{i+1}$
		\EndIf
		\EndFor
	\end{algorithmic}
\end{algorithm}
\begin{algorithm}
	\caption{Inner loop of the adapted truncated Newton method}\label{alg:Inner_Loop}
	\begin{algorithmic}[1]
		%
		%
		\Function{Inner Loop}{}
		\State Set $\ut{r}_1 = -\ut{F}^i$, $\ut{d}_1 = \ut{r}_1$, $\ut{p}_1^i=0$ and $Q_0=0$
		\For{$j=1,2,3,\dots$}
		%
		\If {$\ut{d}_j^T\underline{K}^i\ut{d}_j\le 0$} 
		\If {$j=1$}
		\State \Return $\ut{p}^i=-\ut{F}^i$
		\Else
		\State \Return $\ut{p}^i=\ut{p}_j^i$
		\EndIf
		\EndIf
		\State $\kappa_j=\ut{r}_j^T\ut{r}_j/\ut{d}_j^T\underline{K}^i\ut{d}_j$
		\State $\ut{p}_j^i = \ut{p}_j^i + \kappa_j \ut{d}_j$
		%
		\If {$\sgn{\overline{\ut{B}}\underline{P}\underline{R}\left(\ut{u}_{\mathrm{f}}^i+\ut{p}_j^i\right)} \neq \sgn{\overline{\ut{B}}\underline{P}\underline{R}\ut{u}_{\mathrm{f}}^i}$} 
		\If {$j=1$}
		\State \Return $\ut{p}^i=-\ut{F}^i$
		\Else
		\State \Return $\ut{p}^i=\ut{p}_j^i$
		\EndIf
		\EndIf
		\State $\ut{r}_{j+1} = \ut{r}_j - \kappa_j \underline{K}^i\ut{d}_j$
		\State $Q_j = Q_{j-1} - \frac{1}{2} (\ut{r}_j^T\ut{r}_j)^2/\ut{d}_j^T\underline{K}^i\ut{d}_j$
		\If{$j(Q_j-Q_{j-1})/Q_j \le \eta^i$}
		\State \Return $\ut{p}^i = \ut{p}_j^i$
		\EndIf
		\State $\beta_{j+1} = \ut{r}_{j+1}^T\ut{r}_{j+1}/\ut{r}_j^T\ut{r}_j$
		\State $\ut{d}_{j+1} = \ut{r}_{j+1} + \beta_{j+1} \ut{d}_j$
		\EndFor
		\EndFunction
	\end{algorithmic}
\end{algorithm}
\begin{algorithm}
	\caption{Backtracking Line Search of the truncated Newton Method}\label{alg:Line_Search}
	\begin{algorithmic}[1]
		%
		%
		\Function{Backtracking line search}{}
		\State Set $\alpha = 1$, $\rho\in(0,1)$, $c\in(0,1)$
		\Repeat
		\State $\alpha = \rho \alpha$
		\Until {$\Psi^{\mathrm{h}}(\ut{u}_{\mathrm{f}}^{i}+ \alpha^i\ut{p}^i)\le \Psi^{\mathrm{h}}(\ut{u}_{\mathrm{f}}^i)+c\alpha^i\ut{F}^{\,i^T} \ut{p}^i$}
		\State\Return $\alpha^i=\alpha$		
		\EndFunction
	\end{algorithmic}
\end{algorithm}
%
%
%
%
\section{Comparative performance study}
\label{section:Comparison}
\subsection{Problem statement}
\begin{figure}[htbp]
	\centering
	\includegraphics[width=0.9\linewidth]{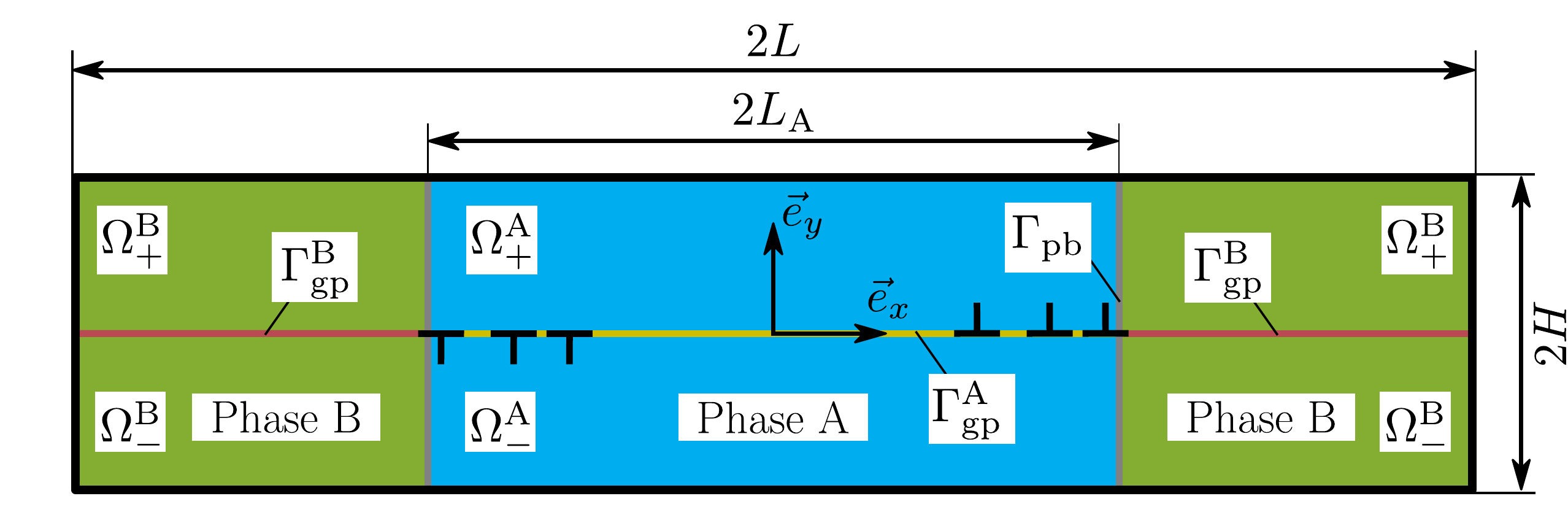}
	\caption{Two-phase microstructure with a single glide plane perpendicular to the phase boundaries.}
	\label{fig:FEM-Model_main_full_simple}
\end{figure}

To demonstrate the stability of the truncated Newton method and the added value of the trust region like extension, the two-phase continuum microstructure illustrated in Figure \ref{fig:FEM-Model_main_full_simple} is considered. It comprises a soft Phase~$\mathrm{A}$ that is flanked by the harder Phase~$\mathrm{B}$. The single glide plane $\Gamma_{gp}^{A/B}$ lies perpendicular to and continuous across the perfectly bonded phase boundary. The local bases of glide plane and phase boundary are thus in alignment with the global basis: $\vec{e}_{t,\mathrm{gp}}=\vec{e}_x$, $\vec{e}_{n,\mathrm{gp}}=\vec{e}_y$ and $\vec{e}_{n,\mathrm{pb}}=\vec{e}_x$. On the outer boundary $\partial\Omega$ a shear deformation is introduced that corresponds to a constant shear stress $\tau=\bar{\tau}t$ for a purely linear elastic medium, i.e., without glide plane. Here, $\bar{\tau}=0.19\cdot(\mu^A b/2\pi d)$ is the target shear load and $t\in[0,1]$ the pseudo-time. Four discrete time points $t_n=n/4$ are considered to simulate the build-up of a four-dislocation pile-up under the incrementally increasing shear deformation. At each time $t_n$, the respective minimisation problem is solved first in sub-increment a. Subsequently, a Peierls-Nabarro dislocation dipole, centred in Phase~$\mathrm{A}$, is introduced and the system is re-equilibrated in sub-increment b. As a result the newly introduced dislocation moves towards the phase boundary and the pile-up process evolves.\par
\begin{figure}[htbp]
	\centering 
	\includegraphics[width=0.9\linewidth]{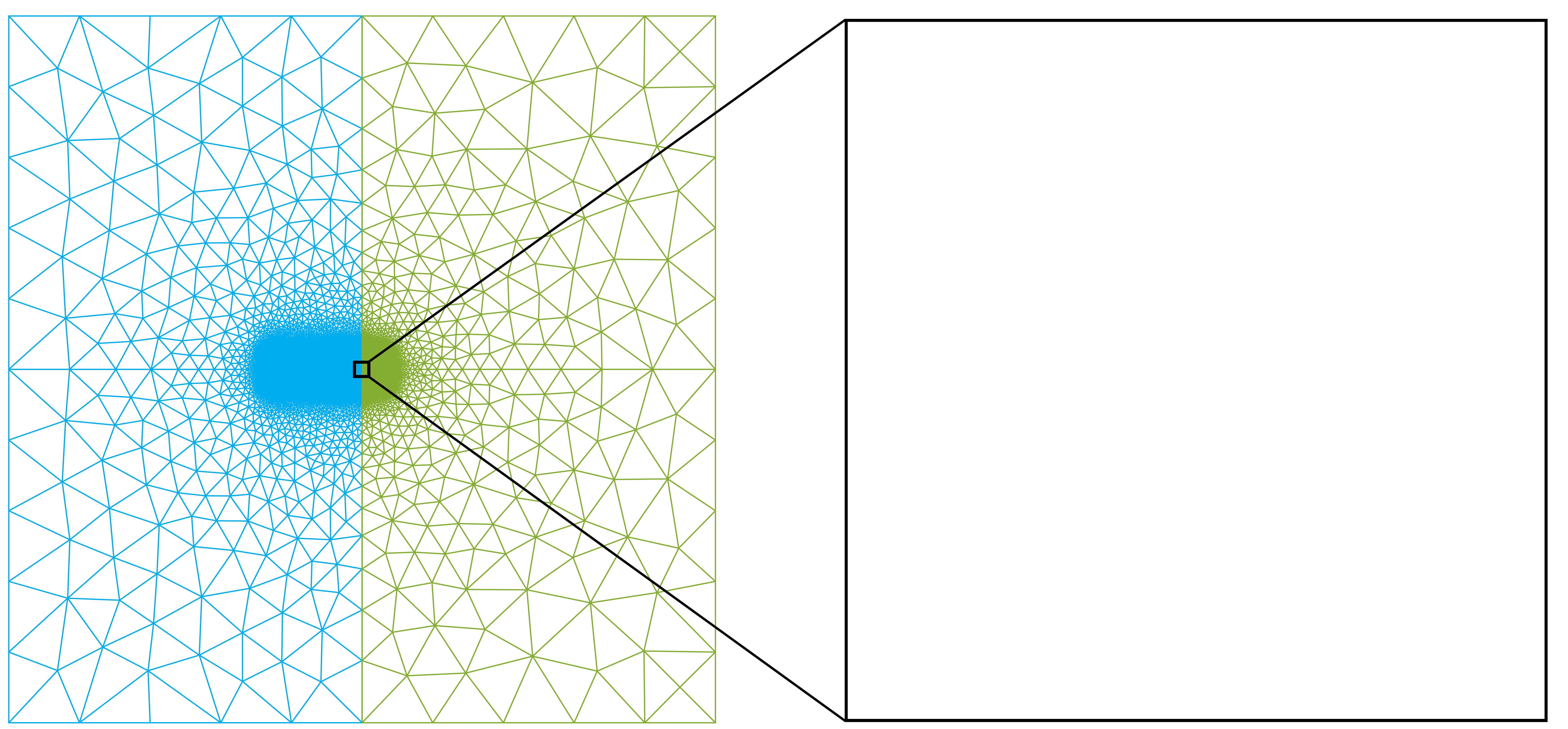}
	\caption{Discretised model for the two-phase microstructure.}
	\label{fig:Meshed_Model}
\end{figure}

Intrinsic to this problem is a symmetry with respect to the model centre ($\vec{x}=\vec{0}$) which is characterised by $\vec{u}(-\vec{x})=-\vec{u}(\vec{x})$. It hence suffices to consider half of the model domain, $0\le x\le L$, and to enforce this symmetry condition along $x=0$ by means of standard FEM tie constraints. For the discretisation linear triangular elements with one central Gauss point for $\Omega_\pm^i$ and linear interface elements with two Gauss points for $\Gamma_{gp}^i$ are used. A convergence study showed a good accuracy in capturing dislocation behaviour for element sizes of $h_{max}=b/8$ in the vicinity of the phase boundary. Away from the region of interest, however, the mesh can be rapidly coarsened. The resulting discretisation is displayed in Figure \ref{fig:Meshed_Model}. It has a total of $482288$ bulk elements and $770$ interface elements. The adopted model parameters are listed in Table \ref{table:Model_parameters}. \par
\begin{table}
	\centering
	\begin{tabular}{|c|l|l|}
		\hline
		Parameters			& Explanation										&	Value  \\\hline
		$2H$ 			& Model height 										& $500b$	\\
		$2L$ 			& Model width 										& $1000b$	\\
		$2L_{\mathrm{A}}$  			& Width of Phase~$\mathrm{A}$									& $500b$	\\
		$h_{max}$		& Maximum element size in refined region			& $b/8$		\\
		$\mu^B/\mu^A$	& Phase contrast									& $1.75$	\\\hline		
	\end{tabular}
	\caption{Model parameters; see Figure \ref{fig:FEM-Model_main_full_simple} for the definition of the geometry parameters}
	\label{table:Model_parameters}
\end{table}
To show the benefits of the truncated Newton method, simulations are carried out with the following numerical solution algorithms: i) the Newton--Raphson method, solved by CG, with line search, ii) the standard truncated Newton method which uses only the conventional truncation criterion \eqref{eq:negative_curvature}, and iii) the adapted truncated Newton method which in addition uses the problem specific truncation criterion \eqref{eq:Disreg_check}. All algorithms have been implemented within an in-house code for the quantitative comparison. The numerical parameters employed are listed in Table \ref{table:Simulation_parameters}.\par
\begin{table}
	\centering
	\begin{tabular}{|c|l|l|}
		\hline
		Parameters			& Description											&	Value	\\\hline
		$\varepsilon_u$	& Displacement tolerance for outer loop					& $1\cdot 10^{-3}$	\\
		$\varepsilon_f$ & Force tolerance for outer loop						& $1\cdot 10^{-3}$	\\
		$\theta$		& Threshold for safeguard of $\eta^i$					& $5\cdot 10^{-2}$	\\
		$\zeta$			& Exponent for safeguard of $\eta^i$					& $1.25$	\\
		$\eta_{\mathrm{l}}$		& Lower bound for tolerance of inner loop				& $5\cdot 10^{-3}$	\\
		$\eta_{\mathrm{u}}$		& Upper bound for tolerance of inner loop				& $1\cdot 10^{-1}$	\\
		$\eta_{\mathrm{ini}}$	& Tolerance of inner loop for first outer iteration		& $1\cdot 10^{-4}$	\\
		$s_{\mathrm{e}}$		& Cut-off radius for additional termination criterion	& $b$	\\
		$\rho$			& Reduction factor for line search						& $0.75$	\\
		$c$				& Gradient factor for Armijo condition					& $1\cdot 10^{-3}$	\\\hline	\end{tabular}
	\caption{Numerical parameters used in the performance assessment}
	\label{table:Simulation_parameters}
\end{table}
\subsection{Simulation results}
\begin{figure}[htbp]
	\centering
	\includegraphics[width=0.95\linewidth]{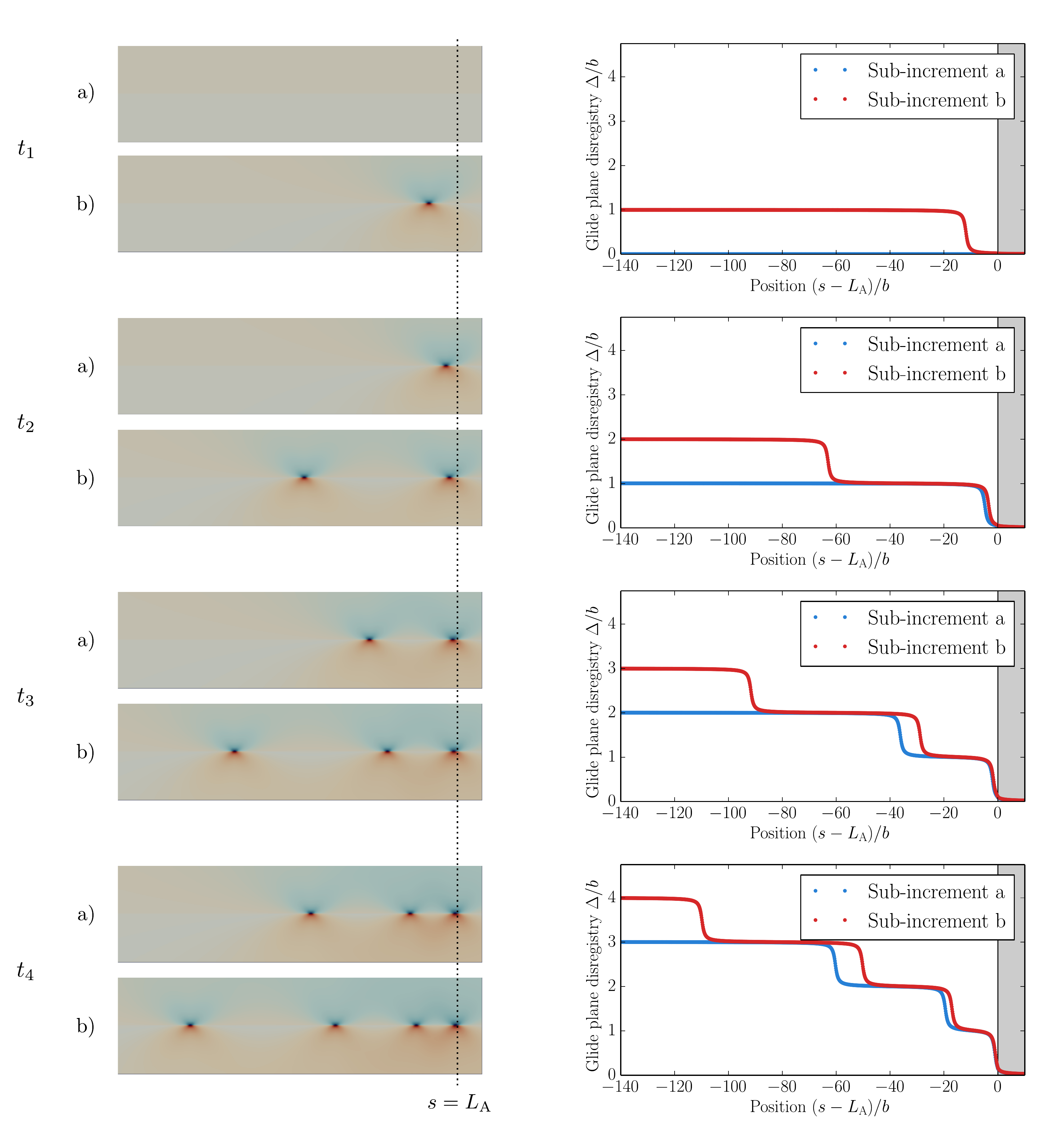}
	\caption{Build-up of a 4-dislocation pile-up at the discrete times $t_n=n/4$ and sub-increments a (before dislocation nucleation) and b (after dislocation nucleation). Left: Close-up view of the qualitative stress-field $\sigma_{xx}$; the dashed line indicates the phase boundary. Right: Disregistry profile along the glide plane; the grey area illustrates the position of Phase~$\mathrm{B}$.}
	\label{fig:Pile-up_evolution}
\end{figure}

The simulation results of the PN-FE model are illustrated in Figure \ref{fig:Pile-up_evolution}. It shows the build-up of the 4-dislocation pile-up at the discrete times $t_n$ and sub-increments a (before dislocation nucleation) and b (after dislocation nucleation). The results illustrate the successive compression of the existing part of the pile-up and the addition of a new dislocation to it with each new level of applied deformation. The left column visualises the dislocation evolution in terms of stress-field $\sigma_{xx}$ for the close-up view $L_{\mathrm{A}}-140b\le x \le L_{\mathrm{A}}+10b$ and $-20b \le y \le 20b$. The dashed line indicates the position of the phase boundary at $x=L_{\mathrm{A}}$. A more detailed insight into the local dislocation structure of the evolving pile-up is given in the right column by the disregistry profile $\Delta$. Here, the grey area indicates the location of Phase~$\mathrm{B}$. Dislocations are located at positions $s$ where $\Delta=(j-1/2)b$, $j=1,2,3,4$.
\par
\subsection{Solver comparison}
\begin{figure}[htbp]
	\centering
	\includegraphics[width=1.\linewidth]{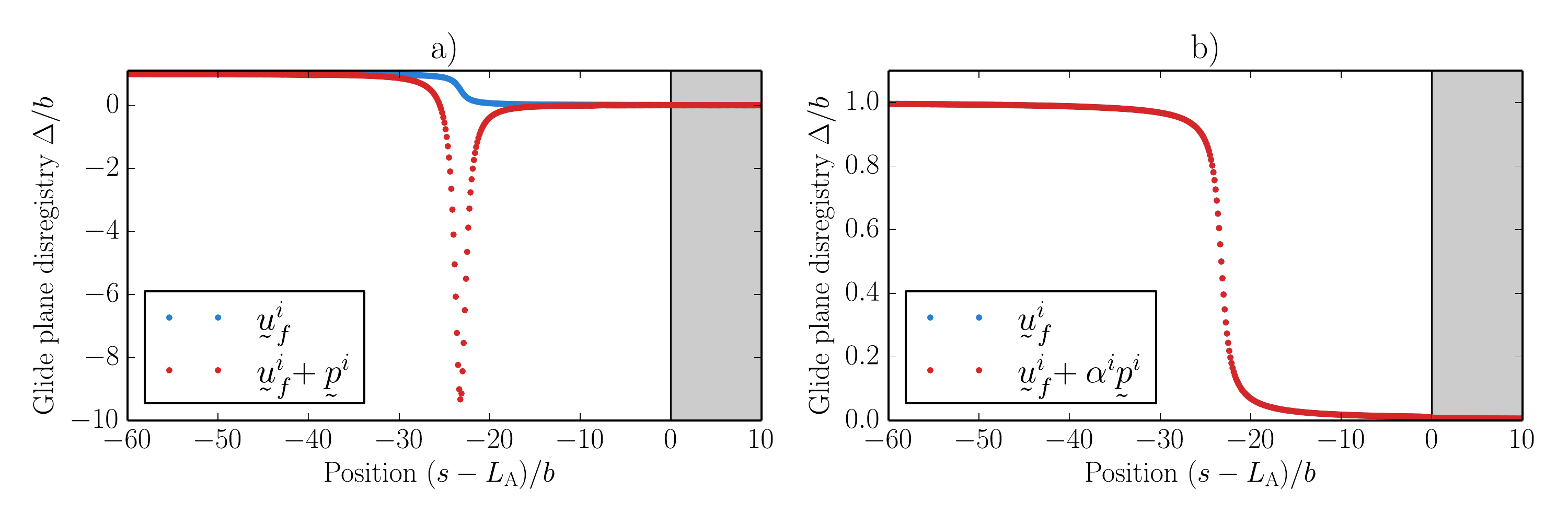}
	\caption{Disregistry profile for the Newton--Raphson method with line search as a result of near singular Hessian $\underline{K}^i$ at time $t_1$ and sub-increment b (after dislocation nucleation) for $i=47$: (a) $\Delta(\protect\ut{u}_{\mathrm{f}}^i+\protect\ut{p}^i)$, i.e., without line search; (b) $\Delta(\protect\ut{u}_{\mathrm{f}}^i+\alpha^i \protect\ut{p}^i)$, i.e., with line search.}
	\label{fig:Disregistry_Profile-CG_fail}
\end{figure}

For the Newton--Raphson method with line search, the missing capability for non-convex minimisation triggers an instability. Convergence is only ensured for sub-increment a at time $t_1$, where no dislocation is present and the problem thus is convex. As soon as a dislocation is introduced (in sub-increment b) the problem is rendered non-convex and the algorithm fails convergence: Starting in outer iteration $i=47$, the non positive definite Hessian leads to a search direction $\ut{p}^i$ for which a sufficient energy decrease requires an extensive line search. As a result, the update $\delta\ut{u}$ is negligibly small and the estimate $\ut{u}^i$ stays in the non-convex region. This is illustrated in Figure \ref{fig:Disregistry_Profile-CG_fail} by the disregistry profiles $\Delta(\ut{u}_{\mathrm{f}}^i+\ut{p}^i)$ (Figure \ref{fig:Disregistry_Profile-CG_fail}a) and $\Delta(\ut{u}_{\mathrm{f}}^i+\alpha^i\ut{p}^i)$ (Figure \ref{fig:Disregistry_Profile-CG_fail}b) in comparison with $\Delta(\ut{u}_{\mathrm{f}}^i)$. Simulations showed that for the conventional Newton--Raphson method (without line search) the solution diverges as a result of the poor search direction, and fails equally in convergence. \par
\begin{figure}[htbp]
	\centering
	\includegraphics[width=0.9\linewidth]{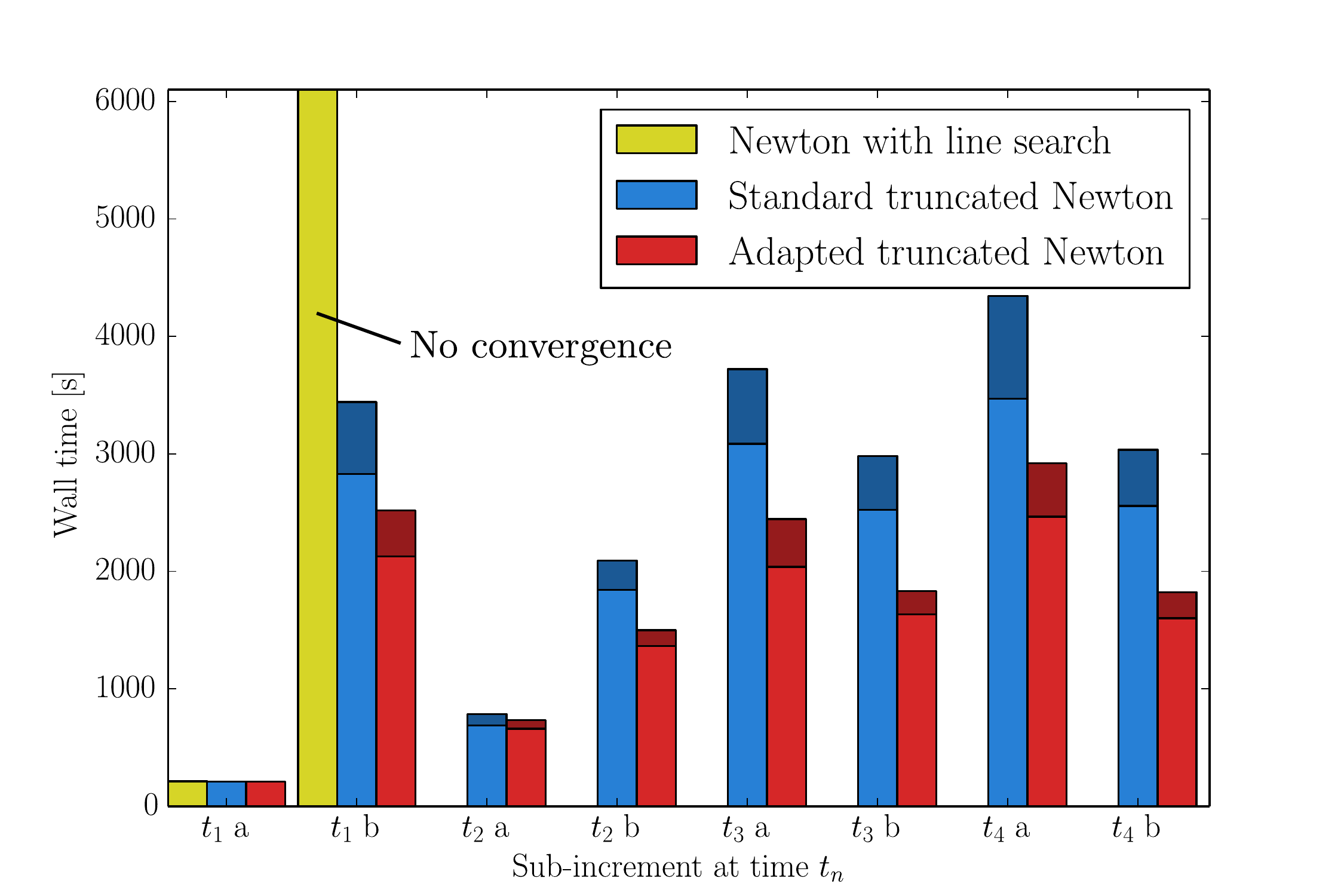}
	\caption{Wall time of the simulation at sub-increments a (before dislocation nucleation) and b (after dislocation nucleation) for each time $t_n$. Comparison between Newton--Raphson method with line search, standard truncated Newton method and adapted truncated Newton method. Light and dark colour represent inner iterations and line search iterations, respectively.}
	\label{fig:Simulation_time}
\end{figure}

Both truncated Newton methods, on the contrary, converge at all times $t_n$. To compare both approaches, at each sub-increment of all $t_n$ the elapsed wall times of the inner iterations and the line search iterations are registered. The results are presented in Figure \ref{fig:Simulation_time}, with the inner iterations and line search iterations represented by the light and dark colour bar, respectively. For each method, the major part of the CPU time is spent in the inner iterations ($\approx 90\%$). 
The comparison of both methods demonstrates the improved performance of the adapted truncated Newton method -- it beats the standard method by roughly a factor of 1.5 in most sub-increments. \par
\begin{figure}[htbp]
	\centering
	\includegraphics[width=1.\linewidth]{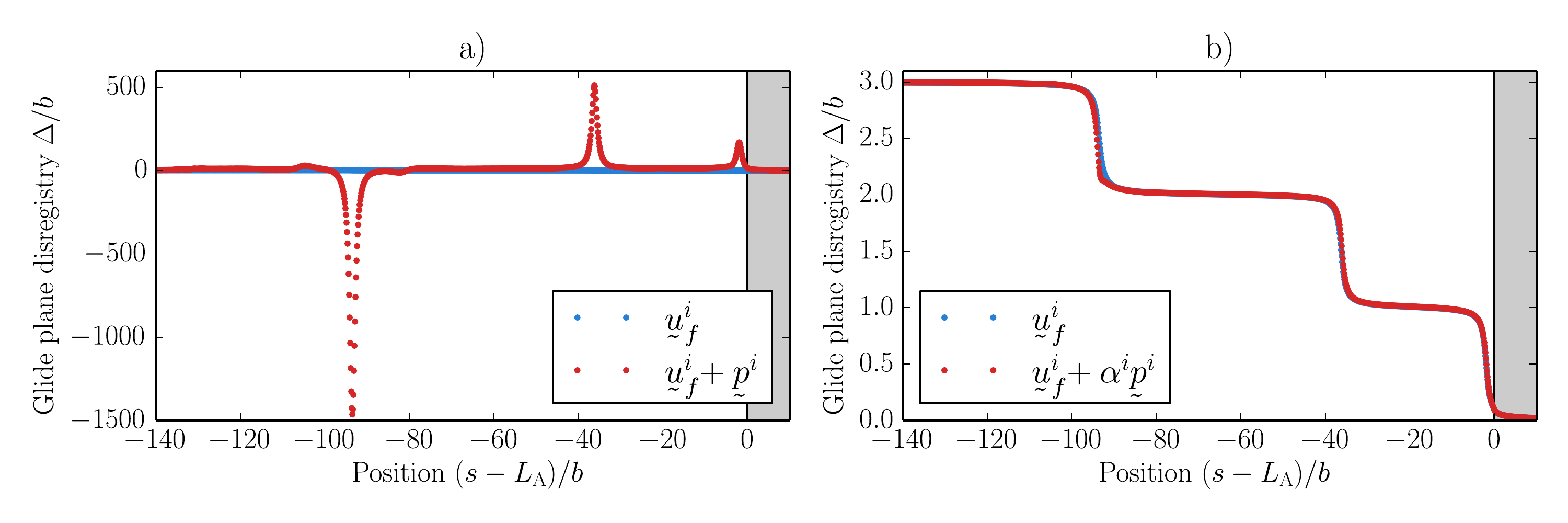}
	\caption{Disregistry profile a) $\Delta(\protect\ut{u}_{\mathrm{f}}^i+\protect\ut{p}^i)$ and b) $\Delta(\protect\ut{u}_{\mathrm{f}}^i+\alpha^i \protect\ut{p}^i)$ for standard truncated Newton as a result of near singular Hessian $\underline{K}^i$ at time $t_3$ and sub-increment b (after dislocation nucleation) for $i=3$.}
	\label{fig:Disregistry_Profile}
\end{figure}

The origin of the differences in wall time lies within the occasional near singularity of the Hessian $\underline{K}^i$ as stated before. Despite the termination due to negative curvature, the near singularity yields sometimes a search direction $\ut{p}^i$ of poor quality. Such an occurrence is illustrated in Figure \ref{fig:Disregistry_Profile}, in terms of the disregistry profiles $\Delta(\ut{u}_{\mathrm{f}}^i+\ut{p}^i)$ in Figure \ref{fig:Disregistry_Profile}a) and $\Delta(\ut{u}_{\mathrm{f}}^i+\alpha^i\ut{p}^i)$ in Figure \ref{fig:Disregistry_Profile}b) in comparison with $\Delta(\ut{u}_{\mathrm{f}}^i)$, in iteration $i=3$ of sub-increment b of time step $t_3$. After acquiring a poor search direction $\ut{p}^i$, the subsequent backtracking line search needs to perform a large number of iterations until the Armijo rule is met and a well-posed disregistry profile with a non-negligible update $\delta\ut{u}$ is recovered. The adapted truncated Newton method, on the contrary, not only prevents an extensive line search, it also reduces the computational expense of the inner loop by avoiding the (inner) iteratively increasing error of $\ut{p}^i$.\par
%
%
%
%
\begin{figure}[htbp]
	\centering
	\includegraphics[width=0.9\linewidth]{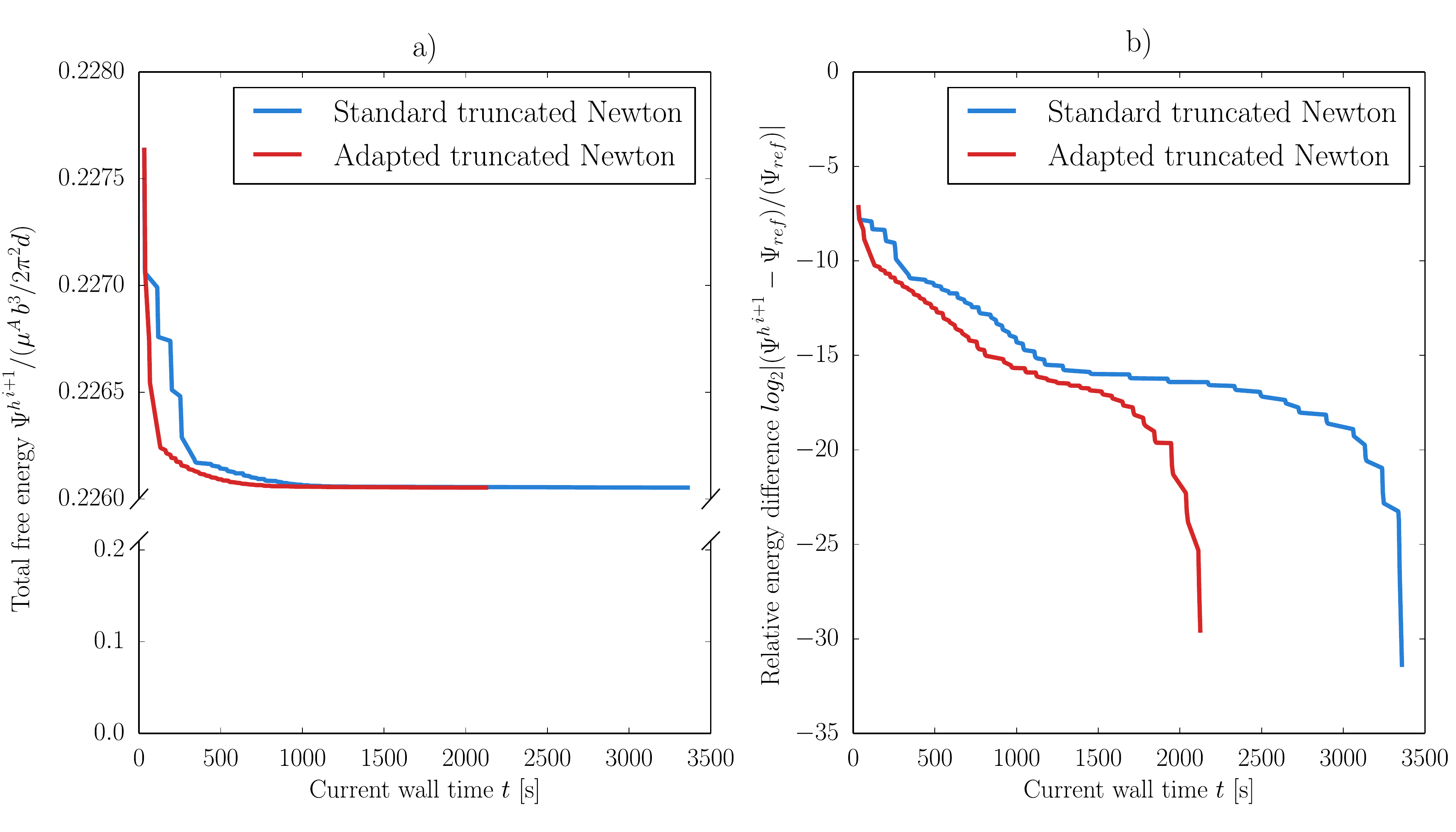}
	\caption{Evolution of the energy of $\Psi^{\mathrm{h}}$ during the iterations in sub-increment b (after dislocation nucleation) at time $t_3$ as a function of the wall time. Comparison between the standard truncated Newton method and the adapted truncated Newton method: a) the normalised iterative energy and b) the relative iterative energy difference with respect to the reference value $\Psi_{ref}$ obtained from a more accurate simulation.}
	\label{fig:Iterative_Energy}
\end{figure}

Such occurrences of poor search directions are accompanied by a slow decrease of the energy $\Psi^{\mathrm{h}}$. This can be shown by the iterative energy $\Psi^i = \Psi^{\mathrm{h}}(\ut{u}_{\mathrm{f}}^i,\ut{u}_{\mathrm{p}})$ as a function of the current wall time. Figure \ref{fig:Iterative_Energy} illustrates the results for sub-increment b (after dislocation nucleation) at time $t_3$ in terms of the normalised quantity $\Psi^{i+1}/(\mu^A b^3/2\pi d)$ in Figure \ref{fig:Iterative_Energy}a) and the relative difference $(\Psi^{i+1}-\Psi_{ref})/\Psi_{ref}$ in Figure \ref{fig:Iterative_Energy}b). $\Psi_{ref}$ is a reference value and represents the equilibrium energy of a refined simulation ($\varepsilon_u=10^{-4}$ and $\varepsilon_f=10^{-6}$). The first plateau observed in the curve for the standard truncated Newton method represents the energy decrease in iteration $i=3$, in which termination due to negative curvature was recorded. Not only the energy decrease is small, but also the computational cost is rather high due to the extensive line search. In the subsequent iterations no negative curvature was recorded. However, occasionally long line searches followed the inner iterations, indicating a poor search direction $\ut{p}^i$. The adapted truncated Newton method, on the contrary, shows an excellent behaviour throughout as a result of the termination prior to the development of a poor $\ut{p}^i$. As a consequence, it reaches convergence about 1.5 times faster than the standard truncated Newton method. \par
It can be concluded that the adapted truncated Newton method provides a stable and efficient numerical scheme for the non-convex minimisation problem of the PN-FE model.
%
%
%
%
\section{Conclusion}
\label{section:Discussion}
With the truncated Newton method an adequate numerical solution algorithm for non-convex energy minimisation was presented. By the example of the PN-FE model it was shown that the standard truncated Newton method already provides a stable numerical scheme for non-convex minimisation where the conventional Newton--Raphson method and the Newton--Raphson method with line search fail. Due to its independence of any regularisation, the truncated Newton method is able to provide results that are not negatively affected by, e.g., damping. However, it has also been shown that its efficiency drops occasionally due to a near singular Hessian. A trust region like adaptation, which is based on the Peierls-Nabarro disregistry profile, alleviates this problem and enhances the efficiency approximately by a factor of 1.5. As the performance of the truncated Newton algorithm strongly depends on the specific dislocation behaviour, it can be anticipated that for more complex problems, i.e., larger dislocation pile-ups, the computational gain increases. Although this addition is limited to the presented PN-FE model, it is believed that in other non-convex minimisation problems the underlying physics may provide a solid basis for similar adaptations. \par
%
%
%
%
\section*{Acknowledgements}
We would like to thank Pratheek Shanthraj of the Max Planck Institute for Iron Research for useful comments on the present work. This research is supported by Tata Steel Europe through the Materials innovation institute (M2i) and Netherlands Organisation for Scientific Research (NWO), under the grant number STW 13358 and M2i project number S22.2.1349a.


\begin{thebibliography}{10}
	\expandafter\ifx\csname url\endcsname\relax
	\def\url#1{\texttt{#1}}\fi
	\expandafter\ifx\csname urlprefix\endcsname\relax\def\urlprefix{URL }\fi
	\expandafter\ifx\csname href\endcsname\relax
	\def\href#1#2{#2} \def\path#1{#1}\fi
	
	\bibitem{Ortiz1999}
	M.~Ortiz, E.~Repetto, {Nonconvex energy minimization and dislocation structures
		in ductile single crystals}, Journal of the Mechanics and Physics of Solids
	47 (1999) 397--462.
	
	\bibitem{Hirth1982}
	J.~Hirth, J.~Lothe, {Theory of Dislocations}, Wiley, New York, 1982.
	
	\bibitem{Joos1994}
	B.~Jo{\'{o}}s, Q.~Ren, M.~Duesbery, {Peierls-Nabarro model of dislocations in
		silicon with generalized stacking-fault restoring forces}, Physical Review B
	50 (1994) 5890--5898.
	
	\bibitem{Schoeck2001}
	G.~Schoeck, {The core structure, recombination energy and Peierls energy for
		dislocations in Al}, Philosophical Magazine A 81 (2001) 1161--1176.
	
	\bibitem{Kochmann2011}
	D.~Kochmann, K.~Hackl, {The evolution of laminates in finite crystal
		plasticity: a variational approach}, Continuum Mechanics and Thermodynamics
	23 (2011) 63--85.
	
	\bibitem{Miehe2004}
	C.~Miehe, M.~Lambrecht, E.~G{\"{u}}rses, {Analysis of material instabilities in
		inelastic solids by incremental energy minimization and relaxation methods:
		Evolving deformation microstructures in finite plasticity}, Journal of the
	Mechanics and Physics of Solids 52 (2004) 2725--2769.
	
	\bibitem{Hackl2008}
	K.~Hackl, D.~Kochmann, {Relaxed potentials and evolution equations for
		inelastic microstructures}, IUTAM Symposium on Theoretical, Computational and
	Modelling Aspects of Inelastic Media 11 (2008) 27--39.
	
	\bibitem{Yalcinkaya2011}
	T.~Yalcinkaya, W.~Brekelmans, M.~Geers, {Deformation patterning driven by rate
		dependent non-convex strain gradient plasticity}, Journal of the Mechanics
	and Physics of Solids 59 (2011) 1--17.
	
	\bibitem{Klusemann2013}
	B.~Klusemann, T.~Yal{\c{c}}inkaya, M.~Geers, B.~Svendsen, {Application of
		non-convex rate dependent gradient plasticity to the modeling and simulation
		of inelastic microstructure development and inhomogeneous material behavior},
	Computational Materials Science 80 (2013) 51--60.
	
	\bibitem{Giessen1995}
	E.~{Van der Giessen}, A.~Needleman, {Discrete dislocation plasticity: a simple
		planar model}, Modelling and Simulation in Materials Science and Engineering
	3 (1995) 689--735.
	
	\bibitem{Vattre2014}
	A.~Vattr{\'{e}}, B.~Devincre, F.~Feyel, R.~Gatti, S.~Groh, O.~Jamond, A.~Roos,
	{Modelling crystal plasticity by 3D dislocation dynamics and the finite
		element method: the discrete-continuous model revisited}, Journal of the
	Mechanics and Physics of Solids 63 (2014) 491--505.
	
	\bibitem{Acharya2001}
	A.~Acharya, {A model of crystal plasticity based on the theory of continuously
		distributed dislocations}, Journal of the Mechanics and Physics of Solids 49
	(2001) 761--784.
	
	\bibitem{Fressengeas2011}
	C.~Fressengeas, V.~Taupin, L.~Capolungo, {An elasto-plastic theory of
		dislocation and disclination fields}, International Journal of Solids and
	Structures 48 (2011) 3499--3509.
	
	\bibitem{Hochrainer2014}
	T.~Hochrainer, S.~Sandfeld, M.~Zaiser, P.~Gumbsch, {Continuum dislocation
		dynamics: towards a physical theory of crystal plasticity}, Journal of the
	Mechanics and Physics of Solids 63 (2014) 167--178.
	
	\bibitem{Wang2010}
	Y.~Wang, J.~Li, {Phase field modeling of defects and deformation}, Acta
	Materialia 58 (2010) 1212--1235.
	
	\bibitem{Mianroodi2016}
	J.~Mianroodi, A.~Hunter, I.~Beyerlein, B.~Svendsen, {Theoretical and
		computational comparison of models for dislocation dissociation and stacking
		fault/core formation in fcc crystals}, Journal of the Mechanics and Physics
	of Solids 95 (2016) 719--741.
	
	\bibitem{Sandfeld2010}
	S.~Sandfeld, T.~Hochrainer, P.~Gumbsch, M.~Zaiser, {Numerical implementation of
		a 3D continuum theory of dislocation dynamics and application to
		micro-bending}, Philosophical Magazine 90 (2010) 3697--3728.
	
	\bibitem{Zhang2015}
	X.~Zhang, A.~Acharya, N.~Walkington, J.~Bielak, {A single theory for some
		quasi-static, supersonic, atomic, and tectonic scale applications of
		dislocations}, Journal of the Mechanics and Physics of Solids 84 (2015)
	145--195.
	
	\bibitem{Bormann2018}
	F.~Bormann, R.~Peerlings, M.~Geers, B.~Svendsen, {A computational approach
		towards modelling dislocation transmission across phase boundaries},
	Submitted, arXiv:1810.08052 [cond-mat.mtrl-sci].
	
	\bibitem{Nash1990}
	S.~Nash, A.~Sofer, {Assessing a search direction within a truncated-Newton
		method}, Operations Research Letters 9 (1990) 219--221.
	
	\bibitem{Xie1999}
	D.~Xie, T.~Schlick, {Efficient implementation of the truncated-Newton algorithm
		for large-scale chemistry applications}, SIAM Journal on Optimization 10
	(1999) 132--154.
	
	\bibitem{Nash2000}
	S.~Nash, {A survey of truncated-Newton methods}, Journal of Computational and
	Applied Mathematics 124 (2000) 45--59.
	
	\bibitem{Fasano2013}
	G.~Fasano, M.~Roma, {Preconditioning Newton-Krylov methods in nonconvex large
		scale optimization}, Computational Optimization and Applications 56 (2013)
	253--290.
	
	\bibitem{Li2001}
	D.-H. Li, M.~Fukushima, {A modified BFGS method and its global convergence in
		nonconvex minimization}, Journal of Computational and Applied Mathematics 129
	(2001) 15--35.
	
	\bibitem{Yuan2018}
	G.~Yuan, Z.~Sheng, B.~Wang, W.~Hu, C.~Li, {The global convergence of a modified
		BFGS method for nonconvex functions}, Journal of Computational and Applied
	Mathematics 327 (2018) 274--294.
	
	\bibitem{Schnabel1999a}
	R.~Schnabel, E.~Eskow, {A revised modified Cholesky factorization algorithm},
	SIAM Journal on Optimization 9 (1999) 1135--1148.
	
	\bibitem{Kelley2013}
	C.~Kelley, L.-Z. Liao, {Explicit pseudo-transient continuation}, Pacific
	Journal of Optimization 9 (2013) 77--91.
	
	\bibitem{Byrd1987}
	R.~Byrd, R.~Schnabel, G.~Shultz, {A trust region algorithm for nonlinearly
		constrained optimization}, SIAM Journal on Numerical Analysis 24 (1987)
	1152--1170.
	
	\bibitem{Conn2000}
	A.~Conn, N.~Gould, P.~Toint, {Trust Region Methods}, SIAM, 2000.
	
	\bibitem{Akaike1959}
	H.~Akaike, {On a successive transformation of probability distribution and its
		application to the analysis of the optimum gradient method}, Annals of the
	Institute of Statistical Mathematics 11 (1959) 1--16.
	
	\bibitem{Wei2008}
	Z.~Wei, G.~Li, L.~Qi, {Global convergence of the Polak-Ribi{\`{e}}re-Polyak
		conjugate gradient method with an Armijo-type inexact line search for
		nonconvex unconstrained optimization problems}, Mathematics of Computation 77
	(2008) 2173--2193.
	
	\bibitem{Yuan2014}
	G.~Yuan, Z.~Wei, Q.~Zhao, {A modified Polak--Ribi{\`{e}}re--Polyak conjugate
		gradient algorithm for large-scale optimization problems}, IIE Transactions
	46 (2014) 397--413.
	
	\bibitem{Sun2015}
	M.~Sun, J.~Liu, {Three modified Polak-Ribi{\`{e}}re-Polyak conjugate gradient
		methods with sufficient descent property}, Journal of Inequalities and
	Applications 2015 (2015) 125.
	
	\bibitem{Nocedal2006}
	J.~Nocedal, S.~Wright, {Numerical Optimization}, Springer Science+Business
	Media, LLC, 2006.
	
	\bibitem{Eisenstat1996}
	S.~Eisenstat, H.~Walker, {Choosing the forcing terms in an inexact Newton
		method}, SIAM Journal on Scientific Computing 17 (1996) 16--32.
	
	\bibitem{Ortega1970}
	J.~M. Ortega, W.~C. Rheinboldt, {9. Rates of convergence-—-general}, in:
	Iterative Solution of Nonlinear Equations in Several Variables, Society for
	Industrial and Applied Mathematics, 2000, pp. 281--298.
	
\end{thebibliography}
\end{document}